\documentclass[12pt]{article}

\usepackage{amsmath,amsfonts,epsfig,mathrsfs,todonotes,yfonts}
\usepackage[driver=pdftex, top=2.5cm, bottom=2.5cm, left=2.5cm, right=2.5cm]{geometry} 
\numberwithin{equation}{section}

\usepackage{hyperref}
 \hypersetup{
     colorlinks=true,
     linkcolor=black,
     filecolor=blue, 
     citecolor=blue,      
     urlcolor=cyan,
     }

\newcommand{\re}[1] {(\ref{#1})}

\newcommand{\ber}{\begin{eqnarray}}
\newcommand{\eer}[1]{\label{#1}\end{eqnarray}}
\newcommand{\eero}{\end{eqnarray}}

\newcommand{\balg}{\begin{align}}
\newcommand{\ealg}{\end{align}}
\newcommand{\bald}{\begin{aligned}}
\newcommand{\eald}{\end{aligned}}

\newcommand{\beq}{\begin{equation}}
\newcommand{\eeq}{\end{equation}}
\newcommand{\bea}{\begin{eqnarray}}
\newcommand{\eea}{\end{eqnarray}}

\newcommand{\nn}{\nonumber}
\newcommand{\na}{\nabla}

\newcommand{\pr}{\prime}
\newcommand{\ppr}{\prime\prime}

\def\HollowBox #1#2{{\dimen0=#1 \advance\dimen0 by -#2
       \dimen1=#1 \advance\dimen1 by #2
        \vrule height #1 depth #2 width #2
        \vrule height 0pt depth #2 width #1
        \llap{\vrule height #1 depth -\dimen0 width \dimen1} 
       \hskip -#2
       \vrule height #1 depth #2 width #2}}
\def\BOX{\HollowBox{.100in}{.010in}}

\newcommand{\auth}{\large Deniz Olgu Devecio\u{g}lu${}^{a}$\footnote{email: dodeve@gmail.com},
Ulf Lindstr\"om${}^{b}$\footnote{email: ulf.lindstrom@physics.uu.se}
and {\"O}zg{\"u}r Sar{\i}o\u{g}lu${}^{c}$\footnote{email: sarioglu@metu.edu.tr (corresponding author)}}

\begin{document}

\begin{flushright}
{\small UUITP-23/24}\\
\vskip 1.5 cm
\end{flushright}

\begin{center}
{\Large{\bf On the maximally symmetric vacua of generic Lovelock gravities}}
\vspace{.75cm}

\auth
\end{center}
\vspace{.5cm}

\centerline{${}^a${\it \small Center for Quantum Spacetime, Sogang University, Seoul, 
121-742, Korea}}
\vspace{.5cm}
\centerline{${}^b${\it \small Department of Physics and Astronomy, Theoretical Physics, 
Uppsala University}}
\centerline{{\it \small SE-751 20 Uppsala, Sweden}}
\centerline{\it and}
\centerline{\it \small Centre for Geometry and Physics, Uppsala University,}
\centerline{\it \small Box 480, SE-751 06 Uppsala, Sweden}
\vspace{.5cm}
\centerline{${}^c${\it \small Department of Physics, Faculty of Arts and Sciences,}}
\centerline{{\it \small Middle East Technical University, 06800, Ankara, T{\"u}rkiye}}

\vspace{1cm}


\centerline{{\bf Abstract}}
\bigskip
\noindent
We survey elementary features of Lovelock gravity and its maximally symmetric vacuum solutions. 
The latter is solely determined by the real roots of a dimension-dependent polynomial. We also 
recover the static spherically symmetric (black hole) solutions of Lovelock gravity using Palais' 
symmetric criticality principle. We show how to linearize the generic field equations of Lovelock 
models about a given maximally symmetric vacuum, which turns out to factorize into the product 
of yet another dimension-dependent polynomial and the linearized Einstein tensor about the 
relevant background. We also describe how to compute conserved charges using linearized field 
equations along with the relevant background Killing isometries. We further describe and discuss
the special vacua which are defined by the simultaneous vanishing of the aforementioned 
polynomials.

\begin{flushright}
\emph{This work is dedicated to the memory of Prof. Stanley Deser.}
\end{flushright}

\bigskip

\small

\renewcommand{\thefootnote}{\arabic{footnote}}
\setcounter{footnote}{0}

\pagebreak
\tableofcontents
\setcounter{page}{2}

\section{Introduction}\label{intro}
General Relativity (GR) has passed numerous observational and experimental tests and 
is essential for our present (classical) understanding of the universe. Its quantum formulation 
is far less developed, although string theory seems to be highly relevant. It is thus of interest 
to investigate extensions and/or modifications of GR to look for novel phenomena and to study 
gravity in various dimensions.

In this context, Lovelock models are of great interest since they offer a natural extension of 
GR to higher dimensions. Lovelock gravities have this flexibility by seamlessly introducing additional 
curvature terms that encode higher-dimensional information, as required by e.g. superstring
theory, without giving up on many of the appealing features of GR. Most notably, they possess
field equations that are strikingly similar to those of GR in four dimensions, yet also have unique 
features in higher dimensions. (See e.g. \cite{Padmanabhan:2013xyr} and the references therein.)

More specifically, Lovelock gravities form a special class of higher curvature gravity 
models since they include at most second order derivatives of the metric tensor and 
they only depend on the Riemann tensor but not its derivatives \cite{Lovelock:1971yv}. 
At the classical level, their Hamiltonian formulation involves only the canonical gravitational 
degrees of freedom of GR \cite{Teitelboim:1987zz}. In a completely classical setting, ``pure 
Lovelock gravity" \cite{Kastor:2012se,Dadhich:2015lra}, the Lovelock model
\( {\cal L} = - 2 \Lambda + \alpha_{[(D-1)/2]} {\cal L}_{[(D-1)/2]} \) 
that is described by the lowest and highest terms in \re{act} and \re{lag}, is claimed to be 
the \emph{unique} generalization of GR to higher \( (D \geq 5) \) dimensions with a 
``kinematic"\footnote{one containing a nontrivial vacuum solution.} static vacuum spacetime 
\cite{Dadhich:2015lra}. At the quantum level, they do not suffer from ghosts that typically 
occur in higher curvature theories \cite{Stelle:1976gc,Zwiebach:1985uq,Zumino:1985dp}. 

Perhaps this is a good place to say a few words as to why we have picked this topic
as a contribution to the special issue dedicated to Stanley Deser's memory. He has been a
driving force in the development of many of the tools used here and he has extensively
utilized these in some of his seminal works that we cite.

The outline of the paper is as follows: In section \ref{love} we briefly describe the key aspects
of Lovelock gravity, its generic action in $D$ dimensions, the source-free field equations
that follow and its maximally symmetric vacua, which are found by determining the real
roots of a polynomial of degree $[(D-1)/2]$. We then give an outline of how the static 
spherically symmetric (black hole) solutions of Lovelock gravity that are already found
in \cite{Wheeler,Whitt:1988ax} can be easily derived using the so-called Weyl-Palais trick
\cite{Weyl,Palais:1979rca} in section \ref{shortcut}. Section \ref{linLove} is devoted to deriving 
the source-free linearized field equations about a given maximally symmetric vacuum, which
in fact factorizes into the product of a polynomial of degree $[(D-1)/2]-1$ and the
linearized Einstein tensor about the given background. The latter polynomial turns out to be
related to the derivative of the former. Section \ref{charge} briefly presents how conserved 
gravitational charges are defined using background Killing isometries for generic Lovelock
theories, which turns out to be proportional to the usual Einstein charges. In section \ref{spev}
we study in detail the \emph{special vacua}, that are defined via the simultaneous vanishing 
of the aforementioned polynomials, about which nontrivial solutions, such as black holes, black 
strings or branes, of a generic Lovelock theory might exist. We then finish up with our
conclusions.

In appendix \ref{appa} we present the explicit forms of the Euler densities for \( D=6, 8, 10 \)
dimensions that are used in Lovelock models. Appendix \ref{appb} gives the explicit expression
of the Euler density \( {\cal L}_{k} \) for a generic \(k \geq 1 \) when the $D$-dimensional metric 
\re{Smet} and its relevant curvature tensors are substituted directly into \( {\cal L}_{k} \).
Appendix \ref{appc} is devoted to the determination of special vacua for the dimensions 
\( 8 \leq D \leq 11 \).

\section{Lovelock gravity}\label{love}
The action of Lovelock theories in $D \geq 3$ spacetime dimensions is given by\footnote{Intrinsically
the treatment here uses the second-order metric formulation and the connection is Levi-Civita. For a
discussion on what happens for the first-order (or Palatini) formulation and other interesting features
of Lovelock theories, see the explicatory review \cite{Padmanabhan:2013xyr}.}
\ber
I = \int d^{D}x \, \sqrt{|g|} \, {\cal L}(g^{ab}, R_{abcd}(g)) \,,
\eer{act}
with the Lagrangian
\ber
{\cal L} := - 2 \Lambda + \sum_{k=1}^{[(D-1)/2]} \alpha_{k}  {\cal L}_{k} \,,
\eer{lag}
where $\Lambda$ is the bare cosmological constant, $[x]$ denotes the integer part of 
a real number $x$, $\alpha_{k}$ are properly chosen coupling constants of appropriate
dimensions with
\ber
{\cal L}_{k} := \frac{1}{2^{k}} \,
\delta^{a_{1} b_{1} \dots a_{k} b_{k}}_{c_{1} d_{1} \dots c_{k} d_{k}} \,
R^{c_{1} d_{1}}{}_{a_{1} b_{1}} \, R^{c_{2} d_{2}}{}_{a_{2} b_{2}} \dots 
R^{c_{k} d_{k}}{}_{a_{k} b_{k}}
\,, \qquad k \geq 1 \,.
\eer{lagk}
Here the generalized Kronecker delta is totally antisymmetric in all up and down indices, and 
is defined as
\ber
\delta^{a_{1} \dots a_{k}}_{c_{1} \dots c_{k}} 
= k! \, \delta^{[a_{1} \dots a_{k}]}_{c_{1} \dots c_{k}}
= k! \, \delta^{a_{1} \dots a_{k}}_{[c_{1} \dots c_{k}]} \,,
\eer{genKro}
and the first few terms\footnote{We give the explicit forms of a few others in the appendix 
\!\ref{appa}.} in the Lagrangian \re{lag} are 
\bea
&& {\cal L}_{1} = R \,, \label{L1} \\
&& {\cal L}_{2} = R^{2} - 4 R_{ab} R^{ab} + R^{cd}{}_{ab} R_{cd}{}^{ab} \,. \label{L2} 
\eea
When $D$ is even, ${\cal L}_{D/2}$ is the Euler density, which gives a topological
invariant when integrated over a compact manifold without a boundary.
The variation of ${\cal L}_{D/2}$ for 
even $D$ is a total derivative and does not contribute to the equations of motion. Moreover 
${\cal L}_{k}$ terms vanish identically for $k > D/2$. So there is a natural difference between
even and odd dimensions: As seen from \re{lag}, the Euler density of the even dimension 
is added each time one moves up to the next odd dimension, whereas no new term is added 
when one moves from an odd dimension to the next even dimension. 

Closely related to Lovelock models is the so-called Chern-Simons (CS) gravity theories
which are defined only in odd dimensions and can be seen as a higher dimensional ($D \geq 5$)
generalization of the CS description of three-dimensional Einstein gravity as a gauge theory.
CS gravities are particular cases of Lovelock models that admit a CS-like action with specific
set of coefficients $\alpha_{k}$ depending on the odd dimension $D$. In complete analogy, there
is also so-called Born-Infeld (BI) gravity theories that exist in even dimensions with a precise
set of coefficients depending on the dimension $D$ 
\cite{Garraffo:2008hu, Arenas-Henriquez:2019rph}.

The source-free field equations of a general Lovelock theory \re{act} are
\ber
E^{a}{}_{b} := \Lambda \delta^{a}_{b} + \sum_{k=1}^{[(D-1)/2]} \alpha_{k} \, 
{\cal G}_{k}{}^{a}{}_{b} = 0 \,, 
\eer{eom}
where
\ber
{\cal G}_{k}{}^{a}{}_{b} := - \frac{1}{2^{k+1}} \,
\delta^{a a_{1} b_{1} \dots a_{k} b_{k}}_{b c_{1} d_{1} \dots c_{k} d_{k}} \,
R^{c_{1} d_{1}}{}_{a_{1} b_{1}} \, R^{c_{2} d_{2}}{}_{a_{2} b_{2}} \dots 
R^{c_{k} d_{k}}{}_{a_{k} b_{k}} \,,
\eer{genG}
and it is easy to see that \( {\cal G}_{1}{}^{a}{}_{b} \) reduces to the ordinary Einstein 
tensor $G^{a}{}_{b}$. It is also apparent that the diffeomorphism invariance of the action 
\re{act} leads directly to a generalized Bianchi identity \( \na_{a} E^{a}{}_{b} = 0 \).

Obviously, Lovelock gravity in $D$ dimensions may have up to 
$[(D-1)/2]$ distinct maximally symmetric vacua $\bar{g}_{ab}$ 
\cite{Boulware:1985wk,Kastor:2015sxa}, for which the Riemann curvature tensor 
is\footnote{We will put a bar on a curvature tensor when working with maximally 
symmetric spacetimes, thus e.g. \( \bar{R}^{cd}{}_{ab} := R^{cd}{}_{ab}(\bar{g}) \)
for the metric $\bar{g}$ of a maximally symmetric spacetime.}
\ber
\bar{R}^{cd}{}_{ab} = \lambda \, \delta^{cd}_{ab} \,, \quad 
\bar{R}^{a}{}_{b} = \lambda \, (D-1) \, \delta^{a}_{b} \,, \quad
\bar{R} = \lambda \, D (D-1) \,, \;\; \mbox{for a real constant} \; \lambda \,,
\eer{Riemcon}
depending on the relevant couplings $\alpha_{k}$. 
The maximally symmetric spacetime \re{Riemcon} is Minkowski when $\lambda=0$,
de Sitter (dS) when $\lambda >0$ and Anti de Sitter (AdS) when $\lambda < 0$.
Since \re{Riemcon} gives
\ber
\bar{{\cal G}}_{k}{}^{a}{}_{b} = - \frac{1}{2} \lambda^{k} \, \frac{(D-1)!}{(D-(2k+1))!} \, \delta^{a}_{b} \,,
\eer{barcalG}
the field equations \re{eom} for a maximally symmetric spacetime \re{Riemcon} imply
\ber
\bar{E}^{a}{}_{b} = 
\Lambda \delta^{a}_{b} + \sum_{k=1}^{[(D-1)/2]} \alpha_{k} \, \bar{{\cal G}}_{k}{}^{a}{}_{b} = 0 \,,
\eer{barE}
that reduces to a polynomial equation (of degree $[(D-1)/2]$) for $\lambda$
\ber
p(\lambda) := \Lambda - \frac{1}{2} \, (D-1)(D-2) \, \sum_{k=1}^{[(D-1)/2]} 
\omega_{k} \, \lambda^{k} = 0 \,.
\eer{polp}
Here we have defined rescaled versions of the parameters $\alpha_{k}$ in the Lagrangian \re{lag}
\ber
\omega_{k} := \alpha_{k} \, \frac{(D-3)!}{(D-(2k+1))!} \,,
\eer{alpome}
for the sake of additional transparency and clarity in what follows, and to rid the ensuing 
calculations of unwieldy numbers. Note in passing that the parameter in front of the usual 
Einstein-Hilbert piece \re{L1} is \( \alpha_{1} = \omega_{1} \) now.

The real roots of the polynomial equation \re{polp} determine physically viable maximally 
symmetric vacua. When $[(D-1)/2]$ is odd, there is at least one maximally symmetric vacua,
but for generic $[(D-1)/2]$ and $\omega_{k}$, there is a wide range of possibilities going from the
``no vacuum case" to ``the maximum number, i.e., $[(D-1)/2]$ distinct vacua case". In general, 
one can, at least in principle, study (hyper)surfaces in the parameter space defined by 
$\omega_{k}$ for which the number of maximally symmetric vacua change \cite{Kastor:2015sxa}.
In fact, a detailed classification of vacuum solutions with spherical, planar or hyperbolic
spatial geometry can be found in \cite{Maeda:2011ii}, where the conditions on the coupling constants
$\omega_{k}$ are also examined to further determine degenerate maximally symmetric vacua.

\section{The shortcut to static black hole solutions}\label{shortcut}
An appealing feature of Lovelock gravity models is that they admit explicit
static spherically symmetric (black hole) solutions that allow for deformations of the 
``Schwarzschild geometry". Even though such solutions are  well-known
\cite{Wheeler,Whitt:1988ax}, here we want to recapitulate an alternative and much shorter 
way for driving this family using the Weyl-Palais trick \cite{Weyl,Palais:1979rca,Deser:2003up} 
of substituting in the Lovelock action \re{act} the gauge-fixed Schwarzschild metrics 
(see \re{Smet} below) endowed with the desired spherical symmetries. This trick 
considerably simplifies the derivation of the relevant field equations and, implicitly upholding 
Bianchi identities, guarantees the determination of all relevant solutions \cite{Palais:1979rca}.

The most general $D$-dimensional static spherically symmetric metric in Schwarzschild
coordinates is\footnote{In fact one may take the metric functions $a$ and $b$ to depend
also on the $t$-coordinate, which, however, drops out of the actions in the ensuing discussion
and can be shown to lead to Birkhoff's theorem \'a la \cite{Deser:2004gi,Deser:2007za}.}
\ber
ds^2 = -a(r) \, b^2(r) \, dt^2 + \frac{dr^2}{a(r)} + r^2 \, d\Omega_{D-2}^{2} \,,
\eer{Smet}
where \( d\Omega_{D-2}^{2} \) is the metric on the unit $(D-2)$-dimensional sphere \( S^{D-2} \).
The essence of the ``symmetric criticality principle" \cite{Palais:1979rca} lies in the
straightforward substitution of \re{Smet} and its relevant curvature tensors (which are 
calculated relatively easily by making use of the ``composite structure" of \re{Smet}) into 
the Lagrangian \re{lag} and its pieces ${\cal L}_{k}$ \re{lagk} that make up the action \re{act}, 
which gives e.g. via \re{L1} and \re{L2}\footnote{In appendix \ref{appb}, we give the 
analogous expressions for the Lagrangians listed in appendix \ref{appa}.}
\begin{align}
{\cal L}_{1} = {}&
- \frac{r \left(b^{\pr} \left(3 r a^{\pr}+2 (D-2) a\right)+2 r a b^{\ppr}\right)+b \left(r^2 a^{\ppr}
+(D-2) \left(2 r a^{\pr}+(D-3) (a-1)\right)\right)}{r^2 b}, \label{L1cal} \\
{\cal L}_{2} = {}& \frac{(D-3) (D-2)}{r^4 b} \Big(
2 r \big\{ b^{\pr} \left[ r (5 a-3) a^{\pr}+2 (D-4) (a-1) a \right] +2 r (a-1) a b^{\ppr} \big\} \Big. \nn \\
& 
+b \big\{ a \left[ 2 r^2 a^{\ppr}-2 (D-4) \left(-2 r a^{\pr}+D-5\right) \right]
+2 r \left[ a^{\pr} \left(r a^{\pr}-2(D-4) \right) -r a^{\ppr} \right] \big.  \nn \\
& \qquad  \qquad \Big. \big. +(D-5) (D-4) (a^2+1)  \big\} \Big) \,, \label{L2cal}
\end{align}
where a prime indicates derivative with respect to the coordinate $r$. 

As a nontrivial and illustrative example, consider the $D=5$ case. Then, the Lovelock action 
\re{act} reduces to the simpler form
\ber
I \to \int_{0}^{\infty} dr \, r^3 \, b \, 
\Big( -2 \Lambda + \alpha_{1}  {\cal L}_{1} + \alpha_{2}  {\cal L}_{2} \Big)  \Big|_{D=5} \,,
\eer{BD5} 
by discarding the contributions from the angular integrations. The reduced field equations that
follow from the calculus of variations on the two metric functions $a(r)$ and $b(r)$ decouple,
and immediately lead to the well-known Boulware-Deser solution \cite{Boulware:1985wk} 
with two separate branches of static black holes \re{Smet} 
\ber
D=5: \qquad b(r) = 1 \,, \qquad  a_{\pm}(r) = 1+ \frac{\alpha_{1} \, r^2}{4 \alpha_{2}} 
\left( 1 \pm \sqrt{1 + \frac{4 \Lambda \alpha_{2}}{3 \alpha_{1}^{2}} + \frac{m}{r^4} } \right) \,,
\eer{GBBH}
for generic values of the parameters \( \alpha_{1}, \alpha_{2} \) and \( \Lambda \), and an
integration constant $m$ that is related to the mass of the black hole(s). 

Obviously, the aforementioned static spherically symmetric (black hole) solutions of Lovelock gravity 
\cite{Wheeler,Whitt:1988ax} can be derived readily in a similar fashion. To this end, the analog 
of \re{BD5} is
\ber
I \to \int_{0}^{\infty} dr \, b \, r^{D-2} \, \Big(  -2 \Lambda + 
\sum_{k=1}^{[(D-1)/2]} \alpha_{k} \, {\cal L}_{k} \Big) \,,
\eer{genLove} 
and one finds, using \re{Lkcal}, that
\begin{align}
b r^{D-2}  {\cal L}_{k} = 
\frac{(D-2)!}{(D-2 k)!} \left( \left( - 2 k a (1-a)^{k-1} b^{\prime} r^{D-2 k} 
+ b \left( r^{D-2 k}  (1-a)^{k} \right)^{\prime} \right)^{\prime} 
- b^{\prime} (1-a)^{k} (r^{D-2 k})^{\prime} \right) \,.
\label{kokgLkcal}
\end{align}
Substituting \re{kokgLkcal} into \re{genLove} and dropping the total derivative terms, the 
relevant reduced action reads
\ber
- \int_{0}^{\infty} dr \left( -2 \Lambda \, b \, r^{D-2} + (D-2) \, b^{\prime} \,
\sum_{k=1}^{[(D-1)/2]}  \, \omega_{k} \,  (1-a)^{k} \, r^{D-2 k-1} \right) \,.
\eer{redact}
after using \re{alpome}.
Once again the variation of \re{redact} with respect to $a(r)$ sets \( b(r) = 1 \), without
loss of generality, whereas the variation of \re{redact} with respect to $b(r)$ gives
\[
2 \Lambda r^{D-2} -  (D-2) \, \sum_{k=1}^{[(D-1)/2]} \, \omega_{k} \, 
\left( (1-a)^{k} \, r^{D-2 k-1} \right)^{\prime} = 0 \,, \]
which is easily integrated to yield
\ber
\sum_{k=1}^{[(D-1)/2]} \, \omega_{k} \, \left( \frac{(1-a)}{r^2} \right)^{k} =
{\cal M} \, r^{1-D} + \frac{2 \Lambda}{(D-1)(D-2)} \,,
\eer{mostgena}
for an integration constant ${\cal M}$. The polynomial for the metric function $a(r)$
\re{mostgena} is precisely the implicit condition on the relevant metric function in 
\cite{Whitt:1988ax} after careful identifications. For example, the Boulware-Deser solution 
\cite{Boulware:1985wk} reproduced in the previous paragraph \re{GBBH} follows from 
setting \( D=5 \) in \re{mostgena}, making use of \re{alpome} and
identifying the parameter $m$ in \re{GBBH} with \( 4 \omega_{2} {\cal M} /  \omega_{1}^{2} \).

\section{Linearized field equations}\label{linLove}
We now turn to the linearization of the source-free field equations of Lovelock models about 
a generic maximally symmetric background. Let us assume that there exists a well-defined 
maximally symmetric vacuum $\bar{g}_{ab}$ for which 
\( E^{a}{}_{b} (\bar{g}) := \bar{E}^{a}{}_{b} = 0 \), i.e. that \( p(\lambda) = 0 \) in \re{polp} 
has at least one real root. Then the source-free field equations \( E^{a}{}_{b} (g) = 0 \) for a more 
general metric $g_{ab}$ can be linearized about this vacuum (or ``background") 
$\bar{g}_{ab}$ using the metric fluctuations (or ``deviations") $h_{ab}$, with 
\( h_{ab} := g_{ab} - \bar{g}_{ab} \), provided that the deviation $h_{ab}$ goes to zero 
``sufficiently fast” as one approaches the background $\bar{g}_{ab}$ that is typically located at 
``the boundary at infinity” for \( \lambda \leq 0 \). Indicating the linearized tensorial quantities 
with a subscript ``L", one finds using \re{Riemcon} that
\ber
& ({\cal G}_{k}{}^{a}{}_{b})_{L} & := - \frac{1}{2^{k+1}} \,
\delta^{a a_{1} b_{1} \dots a_{k} b_{k}}_{b c_{1} d_{1} \dots c_{k} d_{k}} \,
\left( 
(R^{c_{1} d_{1}}{}_{a_{1} b_{1}})_{L} \, \bar{R}^{c_{2} d_{2}}{}_{a_{2} b_{2}} \dots 
\bar{R}^{c_{k} d_{k}}{}_{a_{k} b_{k}}
 \right. \nn  \\
&& \qquad  \left. 
+ \bar{R}^{c_{1} d_{1}}{}_{a_{1} b_{1}} \, (R^{c_{2} d_{2}}{}_{a_{2} b_{2}})_{L} \dots 
\bar{R}^{c_{k} d_{k}}{}_{a_{k} b_{k}}
+ \dots + \bar{R}^{c_{1} d_{1}}{}_{a_{1} b_{1}} \, \bar{R}^{c_{2} d_{2}}{}_{a_{2} b_{2}} \dots 
(R^{c_{k} d_{k}}{}_{a_{k} b_{k}})_{L} 
\right) \,, \nn \\
&& = 
- \frac{k}{2^{k+1}} \,
\delta^{a a_{1} b_{1} \dots a_{k} b_{k}}_{b c_{1} d_{1} \dots c_{k} d_{k}} \,
(R^{c_{1} d_{1}}{}_{a_{1} b_{1}})_{L} \, \bar{R}^{c_{2} d_{2}}{}_{a_{2} b_{2}} \dots 
\bar{R}^{c_{k} d_{k}}{}_{a_{k} b_{k}} \,, \nn \\
&& = 
- \frac{k}{4} \, \frac{(D-3)! \; \lambda^{k-1}}{(D-(2k+1))!} \,
\delta^{a a_{1} b_{1}}_{b c_{1} d_{1}} \, (R^{c_{1} d_{1}}{}_{a_{1} b_{1}})_{L} \,, \nn \\
&& = k \, \frac{(D-3)! \; \lambda^{k-1}}{(D-(2k+1))!} \, (G^{a}{}_{b})_{L} \,,
\eer{calGlin}
where
\ber
( R^{cd}{}_{ab} )_{L} & = & \bar{R}_{abe}{}^{[c} h^{d]e} + 2 \, \bar{\na}_{[a} \bar{\na}^{[d} h_{b]}{}^{c]} 
\,, \\
( R^{a}{}_{b} )_{L} & = & \frac{1}{2} \left( \bar{\na}^{c} \bar{\na}^{a} h_{bc} + 
\bar{\na}_{c} \bar{\na}_{b} h^{ac} - \bar{\na}^{a} \bar{\na}_{b} h - \bar{\BOX} h^{a}{}_{b} \right)
- h^{ac} \bar{R}_{bc} \,, \\
R_{L} & = & {\bar{\na}}_{a} {\bar{\na}}_{b} h^{ab} - \bar{\BOX} h - h^{ab} \bar{R}_{ab} \,, \\
( G^{a}{}_{b} )_{L} & := & ( R^{a}{}_{b} )_{L} - \frac{1}{2} R_{L} \, \delta^{a}_{b} \,.
\eer{Lin}
Here the raising and lowering of all indices are done with respect to the background $\bar{g}_{ab}$,
\( h := \bar{g}^{ab} h_{ab} \), $\bar{\na}$ indicates the covariant derivative with respect to the 
background metric and \( \bar{\BOX} := \bar{\na}_{a} \bar{\na}^{a} \). (See 
e.g. \cite{Lindstrom:2021qrk} and the references therein for details.) With these preliminaries,
it is now straightforward to arrive at the source-free linearized field equations\footnote{This is
again a verification that source-free Lovelock theories indeed represent a natural generalization 
of Einstein's GR with a unitary massless spin-2 field content.}
\ber
(E^{a}{}_{b})_{L} := \sum_{k=1}^{[(D-1)/2]} \, \alpha_{k} \, ({\cal G}_{k}{}^{a}{}_{b})_{L} 
= q(\lambda) \, (G^{a}{}_{b})_{L} = 0 \,,
\eer{lineom}
where we have defined a polynomial $q(\lambda)$ (of degree \( [(D-1)/2] - 1 \)) as
\ber
q(\lambda) :=  \sum_{k=1}^{[(D-1)/2]} k \, \omega_{k} \, \lambda^{k-1} \,. 
\eer{polq}
Taking the trace of \re{lineom}, we find
\ber
\left( 1-\frac{D}{2} \right) \, q(\lambda) \, R_{L} = 0 \,,
\eer{trlineom}
so that \( R_{L} \) has to vanish provided \( q(\lambda) \neq 0 \).

Note in passing that the polynomials \( p(\lambda) \) \re{polp} and \( q(\lambda) \) \re{polq} 
are related to each other as
\ber
\frac{dp}{d\lambda} + \frac{(D-1)(D-2)}{2} \, q(\lambda) = 0 \,.
\eer{dpeqq}

For the case of CS gravities, when the polynomials \( p(\lambda) \) \re{polp} and \( q(\lambda) \) 
\re{polq} simultaneously share at least one real root, say $\tilde{\lambda}$, then the
relevant CS theory about the corresponding maximally symmetric vacuum enjoys a symmetry
enhancement from local Lorentz symmetry to AdS (if \( \tilde{\lambda} < 0 \)) with a different
number of degrees of freedom and modified dynamics. Similar arguments hold for CS dS
(if \( \tilde{\lambda} > 0 \)) and CS Poincaré gravities (if \( \tilde{\lambda} = 0 \)) 
\cite{Arenas-Henriquez:2019rph}.

It is natural to expect that nontrivial solutions, such as black holes, black strings or branes, 
of a generic Lovelock theory about a special vacuum, defined via the simultaneous vanishing
of the polynomials \( p(\lambda) \) \re{polp} and \( q(\lambda) \) \re{polq}, be special themselves. 
At the classical level, they must define peculiar points where the geometric properties of distinct,
e.g. black hole, solution sets merge or bifurcate. It is tempting to speculate that there 
is a holographic interpretation of Lovelock gravity similar to that of ordinary gravity. Then 
these peculiar points should analogously correspond to branching points for the flows of 
(conformal) field theories with different characteristics.

\section{Conserved charges}\label{charge}
A natural question to ask in the context of the linearized theory is about currents and 
conserved charges. The present section contains such a discussion and briefly depicts
an adaption of the covariant generalization of the celebrated ADM mass definition 
\cite{Arnowitt:1962hi} to Lovelock gravity.

Through linearization, the generalized Bianchi identity \( \na_{a} E^{a}{}_{b} = 0 \) can be 
utilized for constructing a conserved vector current 
\ber 
J^{a} := (E^{a}{}_{b})_{L} \, \bar{\xi}^{b} = q(\lambda) \, J_{\rm Ein}^{a} \,, \quad \mbox{where} \quad
J_{\rm Ein}^{a} :=  (G^{a}{}_{b})_{L} \, \bar{\xi}^{b}\,, 
\eer{lincur}
with \( \bar{\na}_{a} J^{a} = 0 \), using a background Killing vector \( \bar{\xi}^{a} \), for which
\( \bar{\na}_{(a} \bar{\xi}_{b)} = 0 \), via similar arguments as for GR \cite{Abbott:1981ff}. Since 
the current \( J_{\rm Ein}^{a} \) naturally leads to a conserved and background gauge invariant 
gravitational charge\footnote{ See e.g. \cite{Abbott:1981ff,Lindstrom:2021qrk,Gunel:2023ypc} 
for details. The charge definition \re{Echar} implicitly assumes
the background to be either Minkowski or AdS, i.e. \( \lambda \leq 0 \), to avoid the
pathologies that arise from the cosmological horizons of dS spaces.}
\ber
Q_{\rm Ein}(\bar{\xi}) := \oint_{\partial \Sigma} d^{D-2}x \, \sqrt{|q|} \, {n}_{[a} \, r_{b]} \, \ell^{ab} \,,
\eer{Echar}
where $\ell_{a b}$ is the potential 2-form of the current 
\( J_{\rm Ein}^{a} := \bar{\na}_{b} \ell^{a b} \), and \( \ell^{a b} = \ell^{[a b]} \) is explicitly
\cite{Abbott:1981ff,Kastor:2004jk,Lindstrom:2021qrk}
\ber
\ell^{a b} (\bar{\xi}) & = & 
\bar{\xi}_{c} \bar{\na}^{[a} h^{b]c} 
+ \bar{\xi}^{[b} \bar{\na}_{c} h^{a]c} 
+ h^{c [b} \bar{\na}_{c} \bar{\xi}^{a]} 
+ \bar{\xi}^{[a} \bar{\na}^{b]} h 
+ \frac{1}{2} h \bar{\na}^{[a} \bar{\xi}^{b]} \,, \\
& = & -3 \, \delta^{abd}_{eci} \, \bar{\xi}^{e} \, \bar{\na}^{i} \, h_{d}{}^{c} 
+ h^{c [b} \bar{\na}_{c} \bar{\xi}^{a]} 
+ \frac{1}{2} h \bar{\na}^{[a} \bar{\xi}^{b]} \,,
\eer{potein}
it immediately follows that the conserved gravitational charge that follows from the current
\( J^{a} \) is
\ber
Q(\bar{\xi}) = q(\lambda) \, Q_{\rm Ein}(\bar{\xi}) \,.
\eer{char}
This means that when \( q(\lambda) = 0 \), all conserved charges \( Q(\bar{\xi}) \) \re{char}, 
including the mass, vanish.

In this work, we are not interested in the thermodynamics of black holes in Lovelock gravity
theories. Here is a number of papers that discuss this issue 
\cite{Cais,Dadhich:2012ma,Ghosh:2020dkk}. (See also the papers that cite these works.)

\section{Solutions at special vacua}\label{spev}
So we see that the real roots of the polynomial equation \( p(\lambda) = 0 \) \re{polp} determine
the physically viable vacua, whereas the polynomial \( q(\lambda) \) \re{polq} is important for
determining the dynamics \re{lineom} and the conserved charges \re{char} of Lovelock theories.
Hence it is natural to ask about the simultaneous real roots of both \( p(\lambda) = 0 \) and 
\( q(\lambda) = 0 \) for each dimension $D \geq 3$. If such a real root exists and is denoted 
by \( \lambda_{r} \), then one can write \( p(\lambda) = (\lambda - \lambda_{r})^2 \, r(\lambda) \),
where the factor \( r(\lambda) \) is a polynomial (of degree \( [(D-1)/2] - 2 \)) since \( p(\lambda) \) 
and \( q(\lambda) \) are related to each other by \re{dpeqq}. That is, the root \( \lambda_{r} \) is
at least a double root of \( p(\lambda) \). In this case the conserved charge \re{char} is no longer
reliable and one has to resort to other charge definitions, see e.g. \cite{Arenas-Henriquez:2019rph}. 

With the ideas presented in the last paragraph of section \ref{linLove} as an extra motivation,
we hereby set out to determine the simultaneous roots of \( p(\lambda) \) \re{polp} and \( q(\lambda) \) 
\re{polq} explicitly to identify the special vacua first. We next find out the static spherically
black hole solutions of Lovelock models about these special vacua. Even though static
spherically symmetric solutions of \( E^{a}{}_{b} = 0 \) \re{eom} in higher dimensions  
with generic $\alpha_{k}$ was originally given in \cite{Wheeler} and later refined in 
\cite{Whitt:1988ax}, the static spherically symmetric black hole solutions\footnote{See 
\cite{Dadhich:2010gu} for the asymptotic large $r$ limit of these black hole solutions, and 
\cite{Dadhich:2012eg} for the behavior of the entropy going as the square of the horizon radius.} 
we explicitly find here are only implicitly contained in these works and do not follow easily. Hence 
we examine carefully some of these special black hole solutions.

\( \bullet \) The simplest examples are the \( D=3 \) and \( D=4 \) cases, for which
\bea
&& D = 3 : \qquad p(\lambda) = \Lambda -  \omega_{1} \lambda = 0 \,, 
\qquad \qquad q(\lambda) = \omega_{1} = 0 \,, \label{pqD3} \\
&& D = 4 : \qquad p(\lambda) = \Lambda - 3 \omega_{1} \lambda = 0 \,, 
\qquad \qquad q(\lambda) = \omega_{1} = 0 \,.  \label{pqD4}
\eea
For either dimension, both \( p(\lambda) = 0 \) and \( q(\lambda) = 0 \) cannot be 
solved nontrivially. The vacuum is simply given by \( \lambda = \Lambda/\omega_{1} \)
when $D=3$, and by \( \lambda = \Lambda/(3 \omega_{1}) \) when $D=4$, 
both with \( \omega_{1} \neq 0 \). 

\( \bullet \) The first non-trivial, but simple and illustrative, example occurs when $D=5$, for which
\ber
D = 5 : \qquad p(\lambda) = \Lambda - 6 \omega_{1} \lambda - 6 \omega_{2} \lambda^{2} 
\,, \qquad \qquad 
q(\lambda) = \omega_{1} + 2 \omega_{2} \lambda = 0 \,.
\eer{pqD5}
The first thing to note is that $p(\lambda)$ has real roots only if 
\( 3 \omega_{1}^{2} + 2 \omega_{2} \Lambda \geq 0 \). As already mentioned and rediscovered
at the end of section \ref{shortcut}, there are two separate branches of static black holes
\re{GBBH} in \(D = 5 \). If one sets the ``mass parameter" $m=0$ in \re{GBBH}, one ends up 
with two maximally symmetric (background) spaces with\footnote{Recall the redefinition \re{alpome}
we made earlier. So \( \alpha_1 = \omega_{1} \) and \( \alpha_{2} = \omega_{2}/2 \) now.}
\ber 
a_{\pm}(r) \Big|_{m=0} = 1 - \lambda_{\pm} r^{2} \,, \qquad 
\lambda_{\pm} = - \frac{\omega_{1}}{2 \omega_{2}} 
\Big( 1 \pm \sqrt{1+ \frac{2 \omega_{2} \Lambda}{3 \omega_{1}^{2}}} \Big) \,, 
\eer{fm=0}
and \( \lambda_{\pm} \) are the two generic roots of the polynomial \( p(\lambda) = 0 \)
\re{pqD5}, provided \( 3 \omega_{1}^{2} + 2 \omega_{2} \Lambda \geq 0 \), as already mentioned.
Note that the two roots \( \lambda_{\pm} \) merge precisely at the common root of 
\( p(\lambda) \) and \( q(\lambda) \), located at \( \bar{\lambda} = - \omega_{1}/(2 \omega_{2}) \), 
provided \( \omega_{2} \neq 0 \) and \( \Lambda =  - 3 \omega_{1}^{2}/(2 \omega_{2}) \). 
On the other hand, if one sets
\( \Lambda =  - 3 \omega_{1}^{2}/(2 \omega_{2}) \) at the very beginning, 
one ends up with two different spacetimes \re{Smet}, for both of which \( b(r) = 1 \), but 
\ber
a_{1} = 1 + \frac{\omega_{1}}{2 \omega_{2}} \, r^{2} \,, \qquad
a_{2} = c + \frac{\omega_{1}}{2 \omega_{2}} \, r^{2} \,, 
\eer{D5a} 
for an arbitrary integration constant $c$. The choice of $a_{1}$ \re{D5a} in \re{Smet} 
clearly corresponds to a maximally symmetric spacetime \re{Riemcon} with 
\( \lambda = - \omega_{1}/(2 \omega_{2}) \). However, $a_{2}$ \re{D5a}
with $c \neq 1$ represents a geometry given by a direct product \( M \times S^{3} \), where
$M$ is a 2-dimensional maximally symmetric spacetime \re{Riemcon} with 
\( \lambda = - \omega_{1}/(2 \omega_{2}) \) and $c \neq 1$ is a scale that sets the radius of 
the sphere $S^{3}$. So in $D=5$, the simultaneous real root of \( p(\lambda) \) and 
\( q(\lambda) \) \re{pqD5} is a bifurcation point in the solution space where a 
maximally symmetric vacuum and a product space \( M \times S^{3} \) emerge together.

\( \bullet \) The discussion for $D=6$ resembles the one for $D=5$. To start with, one now has
\ber
p(\lambda) = \Lambda - 10 \omega_{1} \lambda - 10 \omega_{2} \lambda^{2} = 0 \,, \qquad \qquad
q(\lambda) = \omega_{1} + 2 \omega_{2} \lambda = 0 \,,
\eer{pqD6}
and $p(\lambda)$ has real roots only if \( 5 \omega_{1}^{2} + 2 \omega_{2} \Lambda \geq 0 \).
In analogy to the $D=5$ case, there are again two separate branches of static black 
holes \re{Smet}\footnote{Recall that the static black hole solutions of generic Lovelock models 
in various dimensions can be found in \cite{Whitt:1988ax}.}
\ber
D=6: \qquad b(r) = 1 \,, \qquad  a_{\pm}(r) = 1+ \frac{\omega_{1} \, r^2}{2 \omega_{2}} 
\left( 1 \pm \sqrt{1 + \frac{2 \Lambda \omega_{2}}{5 \omega_{1}^{2}} + \frac{\tilde{m}}{r^5} } \right)  \,,
\eer{D=6BH}
for generic values of the parameters \( \omega_{1}, \omega_{2} \) and \( \Lambda \), and an
integration constant $\tilde{m}$ that is related to the mass of the black hole(s). When one sets the 
parameter $\tilde{m}=0$ in \re{D=6BH}, one again finds two maximally symmetric (background) 
spaces with
\ber 
a_{\pm}(r) \Big|_{\tilde{m}=0} = 1 - \lambda_{\pm} r^{2} \,, \qquad 
\lambda_{\pm} = - \frac{\omega_{1}}{2 \omega_{2}} 
\Big( 1 \pm \sqrt{1+ \frac{2 \omega_{2} \Lambda}{5 \omega_{1}^{2}}} \Big) \,, 
\eer{tildem=0}
where \( \lambda_{\pm} \) are the two generic roots of the polynomial \( p(\lambda) = 0 \)
\re{pqD6}, provided \( 5 \omega_{1}^{2} + 2 \omega_{2} \Lambda \geq 0 \), of course. 
Once again the two roots \( \lambda_{\pm} \) merge at the common root of \( p(\lambda) \) 
and \( q(\lambda) \), located at \( \tilde{\lambda} = - \omega_{1}/(2 \omega_{2}) \), provided 
\( \omega_{2} \neq 0 \) and \( \Lambda =  - 5 \omega_{1}^{2}/(2 \omega_{2}) \). 
However, the discussion is more interesting than the one for $D=5$ when one seeks out 
solutions of the form \re{Smet} by setting \( \Lambda = - 5 \omega_{1}^{2}/(2 \omega_{2}) \) in 
the first place. In that case, apart from the maximally symmetric metric \re{Smet} with \( b(r) = 1 \)
and \( a(r) = 1 - \tilde{\lambda} r^{2} \), there is also the static black hole \re{Smet}, with 
\( b(r) = 1 \) again, but now\footnote{Setting \( \Lambda = - 5 \omega_{1}^{2}/(2 \omega_{2}) \) 
in the solutions of \cite{Whitt:1988ax} is misleading, since it  doesn't recover the two distinct 
classes we find here.} 
\ber 
a(r) = 1 - \tilde{\lambda} \, r^{2} + \frac{\tilde{M}}{\sqrt{r}} \,.
\eer{speca}
Since $q(\lambda) = 0$ precisely for \( \tilde{\lambda} \), this black hole has \emph{vanishing} 
conserved charges as argued in section \!\ref{charge} and one must resort to other means 
to determine e.g. its mass. To this end, a more careful discussion which uses the full
Lovelock action \re{act}, rather than the linearized field equations \( (E^{a}{}_{b})_{L}  \) as
in section \ref{charge}, that also carefully accounts for the boundary terms emerging during 
the variation of the action is needed. However, that in itself is quite a different enterprise 
which we leave aside.

\( \bullet \) The $D=7$ case is much more involved. Now the polynomials read
\ber
p(\lambda) = \Lambda - 15 \omega_{1} \lambda - 15 \omega_{2} \lambda^{2} 
- 15 \omega_{3} \lambda^{3} = 0 \,, \qquad \qquad
q(\lambda) = \omega_{1} + 2 \omega_{2} \lambda + 3 \omega_{3} \lambda^{2} = 0 \,.
\eer{pqD7}
The generic roots of $q(\lambda)$ are
\ber
\lambda_{\pm} 
= \frac{- \omega_{2} \pm \sqrt{\omega_{2}^2 - 3 \omega_{1} \omega_{3}}}{3 \omega_{3}}  \,, 
\eer{rootsq7}
and these are real when \( \omega_3 \leq \omega_2^2 / (3 \omega_1) \).
Each one of these is also a simultaneous root of $p(\lambda)$ if
\ber
\Lambda(\lambda_{\pm}) = \frac{5}{9 \omega_{3}^2} 
\left( 2 \omega_{2}^3 - 9 \omega_{1} \omega_{2} \omega_{3} 
\mp 2 \left( \omega_{2}^2 - 3 \omega_{1} \omega_{3} \right)^{3/2} \right) \,. 
\eer{Lampm7}
Using the generic \( \omega_{k} \, (k = 1,2,3) \) and either a generic $\Lambda$ or one of these 
specially tuned \( \Lambda(\lambda_{\pm}) \) values, one can indeed find black hole solutions 
of the form \re{Smet} (again with \( b(r) = 1 \)) but neither of the corresponding triplets of $a(r)$
are worthy of displaying here given their bulky form.

Sadly, the discussion does not simplify much with \( \omega_{2} = 0 \), 
for which the roots of $q(\lambda)$ read
\ber 
\lambda_{\pm} = \pm \sqrt{-\frac{\omega_{1}}{3 \omega_{3}}} \,, \qquad \mbox{provided} \;\;
\omega_{1}/ \omega_{3} < 0 \,.
\eer{lam7pm}
These are also roots of $p(\lambda)$ if the cosmological constant is tuned to
\ber 
\Lambda(\lambda_{\pm}) = \pm 10 \omega_{1} \sqrt{-\frac{\omega_{1}}{3 \omega_{3}}} \,. 
\eer{spe7Lam}
However, the unwieldiness of the relevant triplets of $a(r)$ persists and we refrain from
displaying them here for the sake of readability.   

\( \bullet \) For completeness sake, we also carefully study the simultaneous real roots of 
the polynomials \( p(\lambda) \) \re{polp} and \( q(\lambda) \) \re{polq} for the dimensions 
\(8 \leq D \leq 11 \). However, the relevant analysis for these cases are rather elaborate so 
we relegate it to appendix \!\ref{appc}. 

\( \bullet \) Finally we would like to examine Lovelock gravities in generic $D \geq 5$
dimensions with the additional requirement that all \( \omega_{k} = 0 \) for \( k>2 \). This 
particular choice coincides with the cosmological Einstein-Gauss-Bonnet theory in 
$D \geq 5$. In that case, $p(\lambda)$ \re{polp} is quadratic and $q(\lambda)$ \re{polq} is 
linear in $\lambda$:
\ber 
D \geq 5: \quad
p(\lambda) = \Lambda 
- \frac{1}{2} (D-1) (D-2) \big( \omega_{1} \lambda + \omega_{2} \lambda^2 \big) \,
\qquad
q(\lambda) =  \omega_{1} + 2 \omega_{2} \lambda \,,
\eer{Spolpq}
and this generalizes the arguments we gave for $D=5$ and $D=6$ above. The polynomial
$p(\lambda)$ has real roots only if \( (D-1) (D-2) \omega_{1}^{2} + 8 \omega_{2} \Lambda \geq 0 \).
The first thing to note is that there are two separate branches of static black holes
\cite{Wiltshire:1988uq}
\ber
D \geq 5: \qquad b(r) = 1 \,, \qquad  a_{\pm}(r) = 1+ \frac{\omega_{1} \, r^2}{2 \omega_{2}} 
\left( 1 \pm 
\sqrt{1 + \frac{8 \Lambda \omega_{2}}{(D-2)(D-1) \omega_{1}^{2}} + \tilde{M} \, r^{1-D}} \right) \,,
\eer{genDBH}
for generic values of the parameters \( \omega_{1}, \omega_{2} \) and \( \Lambda \), and an
integration constant $\tilde{M}$ that is related to the mass of the black hole(s). The two
maximally symmetric (background) spaces follow when $\tilde{M}=0$ in \re{genDBH}
\ber 
a_{\pm}(r) \Big|_{\tilde{M}=0} = 1 - \lambda_{\pm} r^{2} \,, \qquad 
\lambda_{\pm} = - \frac{\omega_{1}}{2 \omega_{2}} 
\Big( 1 \pm \sqrt{1+ \frac{8 \omega_{2} \Lambda}{(D-2)(D-1) \omega_{1}^{2}}} \Big) \,, 
\eer{tilM=0}
provided \( (D-1) (D-2) \omega_{1}^{2} + 8 \omega_{2} \Lambda \geq 0 \). The two roots 
\( \lambda_{\pm} \) merge at \( \tilde{\lambda} = - \omega_{1}/(2 \omega_{2}) \), if
\( \omega_{2} \neq 0 \) and 
\[ \Lambda = - \frac{\omega_{1}^2}{8 \omega_{2}}  (D-1) (D-2) \,. \]
The discussion follows the one we gave for $D=6$ now: Apart from the maximally symmetric 
metric \re{Smet} with \( b(r) = 1 \) and \( a(r) = 1 - \tilde{\lambda} r^{2} \), there is also the static 
black hole \re{Smet} with 
\ber 
D \geq 6: \qquad b(r) = 1 \,, \qquad  
a(r) = 1 - \tilde{\lambda} \, r^{2} + \hat{M} \; r^{(5-D)/2} \,.
\eer{genspeBH}
where $\hat M$ is again an integration constant. Just like its 6-dimensional counterpart 
\re{speca}, since \( q(\tilde{\lambda}) = 0 \),
all conserved charges of this black hole vanish as per the prescription of section \ref{charge}.
This case shows that the special loci of simultaneous real roots of \( p(\lambda) \) \re{polp} 
and \( q(\lambda) \) \re{polq} may provide room for ``massless" black hole solutions that deserve 
to be further studied.

\section{Summary and discussion}\label{conc}
We have studied the basic features of Lovelock gravity, including its generic $D$-dimensional
action, the source-free field equations and its maximally symmetric vacuum solutions. We have
recapitulated that such vacua are determined by the real roots of the polynomial \( p(\lambda) \)
\re{polp} which in general is of degree $[(D-1)/2]$. We have also rederived the generic static 
spherically symmetric black hole solutions of Lovelock gravity using the symmetric criticality 
principle of Palais. We have also sketched how the generic field equations of Lovelock gravity 
can be linearized about a given maximally symmetric vacuum, and shown that the outcome is the
product of the linearized Einstein tensor about the relevant background with a second
polynomial \( q(\lambda) \) \re{polq}, which is of degree $[(D-1)/2]-1$ and is related to the
derivative of \( p(\lambda) \). We have also outlined how conserved gravitational charges can 
be calculated using linearized field equations and background Killing isometries for a generic
Lovelock model. We have shown that the charges are always equal to the product of the 
usual Einstein charges with the polynomial \( q(\lambda) \) \re{polq}. 

We have further studied in detail the determination of the \emph{special vacua}, that are defined as 
the common real roots of the polynomials \( p(\lambda) \) and \( q(\lambda) \). The discussion 
was relatively easy for $3 \leq D\leq 5$. $D=6$ was the simplest interesting case since we 
were able to identify a nontrivial black hole with vanishing conserved charges at the unique 
special vacuum, which has a slightly different nature than the analogous Boulware-Deser-like 
bifurcating black holes already present there. Thus we expect that there exist nontrivial and 
perhaps unusual solutions at those \emph{special vacua} with possibly interesting features 
on the holography side in $D \geq 7$ as well. Even though we were able to analytically
determine the \emph{special vacua} at $D=7$, the corresponding black hole solutions
were hardly encouraging and we have not been able to consolidate our hunch. Unfortunately
the discussion gets even more sharply complicated with $D > 7$. It may pay off to resort to
numerical methods on top of the analytical approach we have employed so far.

\bigskip

\noindent{\bf Acknowledgments}

\noindent
The research of UL was partly supported by the Leverhulme Trust through a Visiting 
Professorship to Imperial College, which is gratefully acknowledged. DOD
was supported by Basic Science Research Program through the National Research 
Foundation of Korea (NRF) funded by the Ministry of Education, Science and Technology 
(2020R1A2C1010372, 2020R1A6A1A03047877).

\bigskip

\appendix

\section{Other terms in the Lagrangian \re{lag}} \label{appa}
Here we list the explicit forms of a few other terms in the Lagrangian \re{lag}:
 \begin{align}
{\cal L}_{3} = {}&16 R_{a}{}^{c} R^{ab} R_{bc}
 - 12 R_{ab} R^{ab} R
 + R^3
 + 24 R^{ab} R^{cd} R_{acbd} 
 + 3 R R_{abcd} R^{abcd}
  - 24 R^{ab} R_{a}{}^{cde} R_{bcde}\nonumber\\
& - 8 R_{a}{}^{e}{}_{c}{}^{f} R^{abcd} R_{bfde}
  + 2 R_{ab}{}^{ef} R^{abcd} R_{cdef} \label{eq:cubic} \,, \\
{\cal L}_{4} = {}&-96 R_{a}{}^{c} R^{ab} R_{b}{}^{d} R_{cd}
 + 48 R_{ab} R^{ab} R_{cd} R^{cd}
 + 64 R_{a}{}^{c} R^{ab} R_{bc} R - 24 R_{ab} R^{ab} R^2
 + R^4 \nonumber\\
& + 96 R^{ab} R^{cd} R R_{acbd}
 + 6 R^2 R_{abcd} R^{abcd} 
 - 96 R^{ab} R R_{a}{}^{cde} R_{bcde}
 - 384 R_{a}{}^{c} R^{ab} R^{de} R_{bdce}\nonumber\\
& + 96 R^{ab} R^{cd} R_{ac}{}^{ef} R_{bdef}
 + 192 R^{ab} R^{cd} R_{a}{}^{e}{}_{c}{}^{f} R_{bedf}
  - 32 R R_{a}{}^{e}{}_{c}{}^{f} R^{abcd} R_{bfde}
 + 8 R R_{ab}{}^{ef} R^{abcd} R_{cdef}\nonumber\\
& + 192 R_{a}{}^{c} R^{ab} R_{b}{}^{def} R_{cdef}
 - 192 R^{ab} R^{cd} R_{a}{}^{e}{}_{b}{}^{f} R_{cedf}
  + 384 R^{ab} R_{a}{}^{cde} R_{b}{}^{f}{}_{d}{}^{h} R_{chef}
  \nonumber\\
& - 24 R_{ab} R^{ab} R_{cdef} R^{cdef} 
 - 96 R^{ab} R_{a}{}^{cde} R_{bc}{}^{fh} R_{defh}
 + 192 R^{ab} R_{a}{}^{c}{}_{b}{}^{d} R_{c}{}^{efh} R_{defh}\nonumber\\
& - 96 R_{a}{}^{e}{}_{c}{}^{f} R^{abcd} R_{b}{}^{h}{}_{e}{}^{i} R_{dhfi}
 - 96 R_{ab}{}^{ef} R^{abcd} R_{c}{}^{h}{}_{e}{}^{i} R_{difh} 
 + 6 R_{ab}{}^{ef} R^{abcd} R_{cd}{}^{hi} R_{efhi} \nonumber\\
& - 48 R_{abc}{}^{e} R^{abcd} R_{d}{}^{fhi} R_{efhi} 
+ 48 R_{a}{}^{e}{}_{c}{}^{f} R^{abcd} R_{b}{}^{h}{}_{d}{}^{i} R_{ehfi}
 + 3 R_{abcd} R^{abcd} R_{efhi} R^{efhi} \label{eq:quartic} \,. 
 \end{align}
 \begin{align}
{\cal L}_{5} = {}&768 R_{a}{}^{c} R^{ab} R_{b}{}^{d} R_{c}{}^{e} R_{de}
 - 640 R_{ab} R^{ab} R_{c}{}^{e} R^{cd} R_{de}
  - 480 R_{a}{}^{c} R^{ab} R_{b}{}^{d} R_{cd} R
 + 240 R_{ab} R^{ab} R_{cd} R^{cd} R \nonumber\\
& + 160 R_{a}{}^{c} R^{ab} R_{bc} R^2
  - 40 R_{ab} R^{ab} R^3
 + R^5
 + 240 R^{ab} R^{cd} R^2 R_{acbd}
 + 10 R^3 R_{abcd} R^{abcd}\nonumber\\
& - 240 R^{ab} R^2 R_{a}{}^{cde} R_{bcde}
 - 1920 R_{a}{}^{c} R^{ab} R^{de} R R_{bdce}\
  + 480 R^{ab} R^{cd} R R_{ac}{}^{ef} R_{bdef}\nonumber\\
& - 1920 R^{ab} R^{cd} R^{ef} R_{ace}{}^{h} R_{bdfh} 
+ 1920 R_{a}{}^{c} R^{ab} R_{d}{}^{f} R^{de} R_{becf}
 + 960 R^{ab} R^{cd} R R_{a}{}^{e}{}_{c}{}^{f} R_{bedf}\nonumber\\
& - 80 R^2 R_{a}{}^{e}{}_{c}{}^{f} R^{abcd} R_{bfde}
 + 20 R^2 R_{ab}{}^{ef} R^{abcd} R_{cdef}
 + 960 R_{a}{}^{c} R^{ab} R R_{b}{}^{def} R_{cdef} \nonumber\\
& + 3840 R_{a}{}^{c} R^{ab} R_{b}{}^{d} R^{ef} R_{cedf}
  - 960 R_{ab} R^{ab} R^{cd} R^{ef} R_{cedf}
 - 960 R^{ab} R^{cd} R R_{a}{}^{e}{}_{b}{}^{f} R_{cedf}\nonumber\\
& - 1920 R_{a}{}^{c} R^{ab} R^{de} R_{bd}{}^{fh} R_{cefh}
 - 3840 R_{a}{}^{c} R^{ab} R^{de} R_{b}{}^{f}{}_{d}{}^{h} R_{cfeh} 
 + 1920 R^{ab} R R_{a}{}^{cde} R_{b}{}^{f}{}_{d}{}^{h} R_{chef}\nonumber\\
& - 120 R_{ab} R^{ab} R R_{cdef} R^{cdef}
 - 3840 R^{ab} R^{cd} R^{ef} R_{acb}{}^{h} R_{defh}
 - 480 R^{ab} R R_{a}{}^{cde} R_{bc}{}^{fh} R_{defh}\nonumber\\
& - 1920 R_{a}{}^{c} R^{ab} R_{b}{}^{d} R_{c}{}^{efh} R_{defh}
 + 960 R_{ab} R^{ab} R^{cd} R_{c}{}^{efh} R_{defh}
 + 960 R^{ab} R R_{a}{}^{c}{}_{b}{}^{d} R_{c}{}^{efh} R_{defh}\nonumber\\
& + 3840 R^{ab} R_{a}{}^{cde} R_{b}{}^{fhi} R_{chf}{}^{j} R_{deij} 
+ 3840 R_{a}{}^{c} R^{ab} R^{de} R_{b}{}^{f}{}_{c}{}^{h} R_{dfeh}
 + 1920 R^{ab} R^{cd} R_{a}{}^{e}{}_{c}{}^{f} R_{be}{}^{hi} R_{dfhi}\nonumber\\
& - 1920 R^{ab} R^{cd} R_{a}{}^{e}{}_{b}{}^{f} R_{ce}{}^{hi} R_{dfhi}
 + 3840 R^{ab} R^{cd} R_{a}{}^{e}{}_{c}{}^{f} R_{b}{}^{h}{}_{f}{}^{i} R_{dhei} 
 - 480 R R_{a}{}^{e}{}_{c}{}^{f} R^{abcd} R_{b}{}^{h}{}_{e}{}^{i} R_{dhfi}\nonumber\\
& - 3840 R^{ab} R^{cd} R_{a}{}^{e}{}_{b}{}^{f} R_{c}{}^{h}{}_{e}{}^{i} R_{dhfi}
  - 3840 R^{ab} R^{cd} R_{ac}{}^{ef} R_{b}{}^{h}{}_{e}{}^{i} R_{difh}
 - 480 R R_{ab}{}^{ef} R^{abcd} R_{c}{}^{h}{}_{e}{}^{i} R_{difh}\nonumber\\
& - 3840 R_{a}{}^{c} R^{ab} R_{b}{}^{def} R_{c}{}^{h}{}_{e}{}^{i} R_{difh}
 + 320 R_{ab} R^{ab} R_{c}{}^{h}{}_{e}{}^{i} R^{cdef} R_{difh} \nonumber\\
&- 768 R_{a}{}^{efh} R^{abcd} R_{bf}{}^{ij} R_{cei}{}^{k} R_{dkhj} 
+ 160 R_{a}{}^{c} R^{ab} R_{bc} R_{defh} R^{defh} 
 + 480 R^{ab} R^{cd} R_{ac}{}^{ef} R_{bd}{}^{hi} R_{efhi} \nonumber\\
& + 30 R R_{ab}{}^{ef} R^{abcd} R_{cd}{}^{hi} R_{efhi} 
 + 960 R_{a}{}^{c} R^{ab} R_{b}{}^{def} R_{cd}{}^{hi} R_{efhi}
 - 80 R_{ab} R^{ab} R_{cd}{}^{hi} R^{cdef} R_{efhi}\nonumber\\
& - 3840 R^{ab} R^{cd} R_{acb}{}^{e} R_{d}{}^{fhi} R_{efhi}
 - 240 R R_{abc}{}^{e} R^{abcd} R_{d}{}^{fhi} R_{efhi} 
 - 1920 R_{a}{}^{c} R^{ab} R_{b}{}^{d}{}_{c}{}^{e} R_{d}{}^{fhi} R_{efhi}\nonumber\\
& + 3840 R^{ab} R_{a}{}^{cde} R_{b}{}^{fhi} R_{chd}{}^{j} R_{efij} 
 + 1920 R^{ab} R_{a}{}^{cde} R_{b}{}^{f}{}_{d}{}^{h} R_{ch}{}^{ij} R_{efij}\nonumber\\
& + 240 R R_{a}{}^{e}{}_{c}{}^{f} R^{abcd} R_{b}{}^{h}{}_{d}{}^{i} R_{ehfi} 
 + 1920 R^{ab} R^{cd} R_{a}{}^{e}{}_{b}{}^{f} R_{c}{}^{h}{}_{d}{}^{i} R_{ehfi}
 - 1920 R^{ab} R^{cd} R_{a}{}^{e}{}_{c}{}^{f} R_{b}{}^{h}{}_{d}{}^{i} R_{eifh}\nonumber\\
& + 3840 R^{ab} R_{a}{}^{cde} R_{b}{}^{f}{}_{d}{}^{h} R_{c}{}^{i}{}_{f}{}^{j} R_{eihj}
 + 1920 R^{ab} R_{a}{}^{cde} R_{bc}{}^{fh} R_{d}{}^{i}{}_{f}{}^{j} R_{ejhi}\nonumber\\
& - 3840 R^{ab} R_{a}{}^{c}{}_{b}{}^{d} R_{c}{}^{efh} R_{d}{}^{i}{}_{f}{}^{j} R_{ejhi}\ 
+ 1920 R_{a}{}^{e}{}_{c}{}^{f} R^{abcd} R_{bf}{}^{hi} R_{d}{}^{j}{}_{h}{}^{k} R_{ekij}\nonumber\\
& + 240 R^{ab} R^{cd} R_{acbd} R_{efhi} R^{efhi} 
 + 15 R R_{abcd} R^{abcd} R_{efhi} R^{efhi}
 + 960 R^{ab} R_{a}{}^{cde} R_{b}{}^{f}{}_{de} R_{c}{}^{hij} R_{fhij}\nonumber\\
& - 480 R^{ab} R_{a}{}^{cde} R_{bc}{}^{fh} R_{de}{}^{ij} R_{fhij}
 + 960 R^{ab} R_{a}{}^{c}{}_{b}{}^{d} R_{c}{}^{efh} R_{de}{}^{ij} R_{fhij}\nonumber\\
& + 1920 R^{ab} R_{a}{}^{cde} R_{bcd}{}^{f} R_{e}{}^{hij} R_{fhij} 
 - 1920 R^{ab} R_{a}{}^{c}{}_{b}{}^{d} R_{c}{}^{e}{}_{d}{}^{f} R_{e}{}^{hij} R_{fhij}\nonumber\\
&  - 480 R_{ab}{}^{ef} R^{abcd} R_{c}{}^{h}{}_{e}{}^{i} R_{di}{}^{jk} R_{fhjk} 
 - 3840 R^{ab} R_{a}{}^{cde} R_{b}{}^{f}{}_{d}{}^{h} R_{c}{}^{i}{}_{e}{}^{j} R_{fihj}\nonumber\\
&+ 960 R^{ab} R_{a}{}^{cde} R_{b}{}^{fhi} R_{c}{}^{j}{}_{de} R_{fjhi}
  - 1920 R_{a}{}^{e}{}_{c}{}^{f} R^{abcd} R_{b}{}^{h}{}_{e}{}^{i} R_{d}{}^{j}{}_{i}{}^{k} R_{fjhk}\nonumber\\
& + 1920 R_{a}{}^{e}{}_{c}{}^{f} R^{abcd} R_{b}{}^{h}{}_{d}{}^{i} R_{e}{}^{j}{}_{i}{}^{k} R_{fjhk} 
- 1920 R_{ab}{}^{ef} R^{abcd} R_{c}{}^{h}{}_{e}{}^{i} R_{d}{}^{j}{}_{h}{}^{k} R_{fjik}\nonumber\\
& - 480 R_{ab}{}^{ef} R^{abcd} R_{cd}{}^{hi} R_{e}{}^{j}{}_{h}{}^{k} R_{fkij} 
+ 1920 R_{abc}{}^{e} R^{abcd} R_{d}{}^{fhi} R_{e}{}^{j}{}_{h}{}^{k} R_{fkij}\nonumber\\
& - 80 R_{abcd} R^{abcd} R_{e}{}^{j}{}_{h}{}^{k} R^{efhi} R_{fkij}
 - 240 R^{ab} R_{a}{}^{cde} R_{bcde} R_{fhij} R^{fhij}\nonumber\\
& + 24 R_{ab}{}^{ef} R^{abcd} R_{cd}{}^{hi} R_{ef}{}^{jk} R_{hijk}
 - 480 R_{abc}{}^{e} R^{abcd} R_{d}{}^{fhi} R_{ef}{}^{jk} R_{hijk}\nonumber\\
& + 20 R_{abcd} R^{abcd} R_{ef}{}^{jk} R^{efhi} R_{hijk}
 + 480 R_{abc}{}^{e} R^{abcd} R_{d}{}^{f}{}_{e}{}^{h} R_{f}{}^{ijk} R_{hijk}\nonumber\\
& + 960 R_{ab}{}^{ef} R^{abcd} R_{c}{}^{h}{}_{e}{}^{i} R_{d}{}^{j}{}_{f}{}^{k} R_{hjik}
- 384 R_{a}{}^{e}{}_{c}{}^{f} R^{abcd} R_{b}{}^{h}{}_{d}{}^{i} R_{e}{}^{j}{}_{f}{}^{k} R_{hkij} \,.\label{eq:quintic}
\end{align}

\section{Explicit form of \texorpdfstring{${\cal L}_{k\geq 1}$}{} for 
the metric \re{Smet}} \label{appb}
Here we first give the explicit expressions of the Lagrangians listed in appendix \ref{appa}
when \re{Smet} and its relevant curvature tensors are substituted into them:
 \begin{align}
{\cal L}_{3} = {}& \frac{(D-2)! \, (1-a)}{(D-6)! \, r^6 \, b} \Big(
3 r \big\{ b^{\pr} \left[ r (7 a-3) a^{\pr}+2 (D-6) (a-1) a\right] +2 r (a-1) a b^{\ppr} \big\} \Big. \nn \\
& 
+b \big\{ a \left[ 3 r^2 a^{\ppr}-2 (D-6) \left(-3 r a^{\pr}+D-7 \right) \right]
+3 r \left[ a^{\pr} \left(2 r a^{\pr}-2(D-6) \right) -r a^{\ppr} \right] \big.  \nn \\
& \qquad  \qquad \Big. \big. +(D-7) (D-6) (a^2+1)  \big\} \Big)
\label{L3cal} \,, \\
{\cal L}_{4} = {}& \frac{(D-2)! \, (1-a)^2}{(D-8)! \, r^8 \, b} \Big(
4 r \big\{ b^{\pr} \left[ r (9 a-3) a^{\pr}+2 (D-8) (a-1) a \right] +2 r (a-1) a b^{\ppr} \big\} \Big. \nn \\
& 
+b \big\{ a \left[ 4 r^2 a^{\ppr}-2 (D-8) \left(-4 r a^{\pr}+D-9 \right) \right]
+4 r \left[ a^{\pr} \left(3 r a^{\pr}-2(D-8) \right) -r a^{\ppr} \right] \big.  \nn \\
& \qquad  \qquad \Big. \big. +(D-9) (D-8) (a^2+1)  \big\} \Big)
\label{L4cal} \,,  \\
{\cal L}_{5} = {}&  \frac{(D-2)! \, (1-a)^3}{(D-10)! \, r^{10} \, b} \Big(
5 r \big\{ b^{\pr} \left[ r (11 a-3) a^{\pr}+2 (D-10) (a-1) a \right] +2 r (a-1) a b^{\ppr} \big\} \Big. \nn \\
& 
+b \big\{ a \left[ 5 r^2 a^{\ppr}-2 (D-10) \left(-5 r a^{\pr}+D-11\right) \right]
+5 r \left[ a^{\pr} \left(4 r a^{\pr}-2(D-10) \right) -r a^{\ppr} \right] \big.  \nn \\
& \qquad  \qquad \Big. \big. +(D-11) (D-10) (a^2+1)  \big\} \Big) \label{L5cal} \, .
\end{align}
A quick glance at \re{L2cal}, \re{L3cal}, \re{L4cal} and \re{L5cal} reveals the structure of 
generic\footnote{It is not obvious at all but in fact 
\re{Lkcal} also reduces to \re{L1cal} when $k=1$, hence the subscript $k \geq 1$.}
${\cal L}_{k}$ for the $D$-dimensional metric \re{Smet}:
 \begin{align}
{\cal L}_{k \geq 1} = {}& \frac{(D-2)! \, (1-a)^{k-2}}{(D-2 k)! \, r^{2 k} \, b} \Big(
k r \big\{ b^{\pr} \left[ r ((2 k+1) a-3) a^{\pr}+2 (D-2 k) (a-1) a \right] +2 r (a-1) a b^{\ppr} \big\} \Big. \nn \\
&
+b \big\{ a \left[ k r^2 a^{\ppr}-2 (D-2 k) \left(-k r a^{\pr}+D-2 k -1 \right) \right]
+k r \left[ a^{\pr} \left((k-1) r a^{\pr}-2(D-2k) \right) -r a^{\ppr} \right] \big. \nn \\
& \qquad  \qquad \Big. \big. +(D-2k-1) (D-2k) (a^2+1)  \big\} \Big) \label{Lkcal} \, .
\end{align}

\section{Simultaneous roots of \texorpdfstring{$p(\lambda)$}{} \re{polp} and 
\texorpdfstring{$q(\lambda)$}{} \re{polq}} \label{appc}
In this appendix we carefully study the simultaneous real roots of \( p(\lambda) \) \re{polp} 
and \( q(\lambda) \) \re{polq} for the dimensions \(8 \leq D \leq 11 \), given a Lovelock theory 
\re{lag} (with \re{lagk}) containing the parameters $\Lambda$ and $\alpha_{k}$, or equivalently 
$\omega_{k}$ \re{alpome}, \((1 \leq k \leq [(D-1)/2])\). Our strategy is to first figure out the 
conditions that guarantee the reality of the roots of the smaller-degree polynomial \( q(\lambda) \), 
assuming that the coefficient \( \alpha_{[(D-1)/2]} \neq 0 \), and to later tune up the bare cosmological 
constant $\Lambda$ so that the real roots of \( q(\lambda) \) are also roots of \( p(\lambda) \). 
Throughout we also assume that \( \alpha_{1} \), or identically \( \omega_{1} \), is always 
strictly positive, i.e. $\omega_{1}>0$.

After the discussion of the most general case, we also examine the simpler example of the 
vanishing of all $\alpha_{k}$, hence $\omega_{k}$ \re{alpome}, except for $\omega_{1}$ and 
the highest relevant $k$, i.e., the case with \( \omega_{1} > 0 \), \( \omega_{[(D-1)/2]} \neq 0 \) 
and \( \omega_{k} = 0 \) for \( 1 < k < [(D-1)/2] \) for the relevant dimensions \( 8 \leq D \leq 11 \). 
Note that the latter model corresponds to the cosmological Einstein theory coupled with the 
highest order Euler density for \( 8 \leq D \leq 11 \). 

\subsection{\texorpdfstring{$D=8$}{}}\label{sD8}
The relevant polynomials are 
\ber
p(\lambda) = \Lambda - 21 \omega_{1} \lambda - 21 \omega_{2} \lambda^{2} 
- 21 \omega_{3} \lambda^{3} = 0 \,, \qquad \qquad
q(\lambda) = \omega_{1} + 2 \omega_{2} \lambda + 3 \omega_{3} \lambda^{2} = 0 \,,
\eer{pqD8}
with the same $q(\lambda)$ \re{pqD7} as in $D=7$. The generic roots of $q(\lambda)$ are
identical to those \re{rootsq7} of the $D=7$ case, which are real for 
\( \omega_3 \leq \omega_2^2 / (3 \omega_1) \).
These are also simultaneous roots of $p(\lambda)$ provided
\[ \Lambda(\lambda_{\pm}) = \frac{7}{9 \omega_{3}^2} 
\left( 2 \omega_{2}^3 - 9 \omega_{1} \omega_{2} \omega_{3}
\mp 2 \left( \omega_{2}^2 - 3 \omega_{1} \omega_{3} \right)^{3/2} \right) \,, \]
7/5 times those \re{Lampm7} of the $D=7$ case. The analogous discussion for what happens 
when \( \omega_{2} = 0 \) leads to $\lambda_{\pm}$ identical to those \re{lam7pm} of the 
$D=7$ case again. These are also roots of  $p(\lambda)$ if 
\[ \Lambda(\lambda_{\pm}) = \pm 14 \omega_{1} \sqrt{-\frac{\omega_{1}}{3 \omega_{3}}} \,, \]
which is again 7/5 times those \re{spe7Lam} of the $D=7$ case. 

\subsection{\texorpdfstring{$D=9$}{}}\label{sD9}
Now
\ber
&& p(\lambda) = \Lambda - 28 \omega_{1} \lambda - 28 \omega_{2} \lambda^{2} 
- 28 \omega_{3} \lambda^{3} - 28 \omega_{4} \lambda^{4} = 0 \,, \nn \\
&& q(\lambda) = \omega_{1} + 2 \omega_{2} \lambda + 3 \omega_{3} \lambda^{2} 
+ 4 \omega_{4} \lambda^{3}= 0 \,.
\eer{pqD9}
The procedure for determining the roots of the cubic polynomial $q(\lambda)$ \re{pqD9} is 
a well-known, e.g. \cite{abrasteg}, but onerous task. Let us keep the discussion succinct 
and state the most relevant points. First define the following auxiliary variables:
\begin{eqnarray*}
&& Q := \frac{\omega_{2}}{6 \omega_{4}} - \frac{\omega_{3}^{2}}{16 \omega_{4}^{2}} \,,  \\
&& P := - \frac{\omega_{1}}{8 \omega_{4}} + \frac{\omega_{2} \omega_{3}}{16 \omega_{4}^{2}}
- \frac{\omega_{3}^{3}}{64 \omega_{4}^{3}} \,, \\
&& \Delta := Q^3 + P^2 = \frac{\omega_{1}^{2}}{64 \omega_{4}^{2}}
- \frac{\omega_{1} \omega_{2} \omega_{3}}{64 \omega_{4}^{3}}
+ \frac{\omega_{1} \omega_{3}^{3}}{256 \omega_{4}^{4}}
+ \frac{\omega_{2}^{3}}{216 \omega_{4}^{3}}
- \frac{\omega_{2}^{2} \omega_{3}^{2}}{768 \omega_{4}^{4}} \,. \\
&& \Xi := \sqrt[3]{P+\sqrt{\Delta}} \,, \qquad \qquad \Upsilon := \sqrt[3]{P-\sqrt{\Delta}} \,. 
\end{eqnarray*}
Then the \emph{formal roots} of the polynomial $q(\lambda)$ \re{pqD9} are given by
\begin{eqnarray*}
&& \lambda_{1} = - \frac{\omega_{3}}{4 \omega_{4}} + (\Xi + \Upsilon) \,, \\
&& \lambda_{2} = - \frac{\omega_{3}}{4 \omega_{4}} - \frac{1}{2} (\Xi + \Upsilon) + \mathrm{i} 
\frac{\sqrt{3}}{2} (\Xi - \Upsilon) \,, \\
&& \lambda_{3} = - \frac{\omega_{3}}{4 \omega_{4}} - \frac{1}{2} (\Xi + \Upsilon) - \mathrm{i} 
\frac{\sqrt{3}}{2} (\Xi - \Upsilon) \,, 
\end{eqnarray*}
and it follows \cite{abrasteg} that when \\ 
i) $\Delta>0$, there exist one real and a pair of complex conjugate roots of $q(\lambda)$ \re{pqD9}; \\
ii) $\Delta=0$, all roots of $q(\lambda)$ \re{pqD9} are real and at least two of them are equal; \\
iii) $\Delta<0$, all roots of  $q(\lambda)$ \re{pqD9} are real and unequal. \\
Depending on each separate case, the \emph{real roots} \( \lambda_{i}, \, (i=1,2,3), \) 
can be used in the polynomial $p(\lambda)$ \re{pqD9} to tune up the bare cosmological 
constant $\Lambda$.

The analysis simplifies considerably when one sets \(\omega_{2} =  \omega_{3} = 0 \), for which
\[ Q = 0 \,, \qquad P = - \frac{\omega_{1}}{8 \omega_{4}} \,, \qquad \Delta = P^2 > 0 \,, \qquad
\Xi = \sqrt[3]{2 P} \,, \qquad \Upsilon = 0 \,, \]
and there is only one real root of $q(\lambda)$ at
\ber 
\lambda_{r} = \sqrt[3]{- \frac{\omega_{1}}{4 \omega_{4}}} \,, 
\eer{spelamr}
regardless of the sign of \( \omega_{4} \). This is also a root of $p(\lambda)$ if 
\ber 
\Lambda(\lambda_{r}) = 21 \omega_{1} \sqrt[3]{-\frac{\omega_{1}}{4 \omega_{4}}} \,. 
\eer{Lamfor9}

\subsection{\texorpdfstring{$D=10$}{}}\label{sD10}
Now
\ber
p(\lambda) = \Lambda - 36 \omega_{1} \lambda - 36 \omega_{2} \lambda^{2} 
- 36 \omega_{3} \lambda^{3} - 36 \omega_{4} \lambda^{4} = 0 \,,
\eer{pqDten}
and $q(\lambda)$ is identical to the one \re{pqD9} in $D=9$, so the discussion closely follows 
that of the $D=9$ case, with identical expressions for $Q$, $P$, $\Delta$, $\Xi$, $\Upsilon$
and the formal roots \( \lambda_{i}, \, (i=1,2,3), \) as in the previous subsection \!\ref{sD9}. 
Once again the \emph{real} $\lambda_{i}$ can be used in the polynomial $p(\lambda)$ 
\re{pqDten} to tune up the bare cosmological constant $\Lambda$.

The analogous discussion for what happens when one sets \( \omega_{2} = \omega_{3} = 0 \) 
is also identical to the one in the $D=9$ case, having the same special root \( \lambda_{r} \) 
\re{spelamr} as in $D=9$, with only the specially tuned-up $\Lambda$ given by
\ber 
\Lambda(\lambda_{r}) = 27 \omega_{1} \sqrt[3]{-\frac{\omega_{1}}{4 \omega_{4}}}  
\eer{lamforten}
instead of \re{Lamfor9}.

\subsection{\texorpdfstring{$D=11$}{}}\label{eleven}
Finally 
\ber
&& p(\lambda) = \Lambda - 45 \omega_{1} \lambda - 45 \omega_{2} \lambda^{2} 
- 45 \omega_{3} \lambda^{3} - 45 \omega_{4} \lambda^{4} - 45 \omega_{5} \lambda^{5} = 0  \,, \nn \\
&& q(\lambda) = \omega_{1} + 2 \omega_{2} \lambda + 3 \omega_{3} \lambda^{2} 
+ 4 \omega_{4} \lambda^{3}+ 5 \omega_{5} \lambda^{4} = 0 \,.
\eer{pqDele}
It is important to understand the nature of the roots of $q(\lambda)$ \re{pqDele}. To this
end, we follow the discussion in \cite{Wiki} and first define the following parameters with
rather unwieldy coefficients
\begin{eqnarray*}
&& \Delta := 16 \left( 2000 \omega_{1}^3 \omega_{5}^3
- 2400 \omega_{1}^2 \omega_{2} \omega_{4} \omega_{5}^2 
- 1800 \omega_{1}^2 \omega_{3}^2 \omega_{5}^2
+ 2160 \omega_{1}^2 \omega_{3} \omega_{4}^2 \omega_{5} 
- 432 \omega_{1}^2 \omega_{4}^4 
\right. \\
&& \quad \quad 
+ 2700 \omega_{1} \omega_{2}^2 \omega_{3} \omega_{5}^2
- 120 \omega_{1} \omega_{2}^2 \omega_{4}^2 \omega_{5}
- 1800 \omega_{1} \omega_{2} \omega_{3}^2 \omega_{4} \omega_{5}
+ 432 \omega_{1} \omega_{2} \omega_{3} \omega_{4}^3 
+ 405 \omega_{1} \omega_{3}^4 \omega_{5} \\
&& \quad \quad \left.
- 108 \omega_{1} \omega_{3}^3 \omega_{4}^2
- 675 \omega_{2}^4 \omega_{5}^2 
+ 540 \omega_{2}^3 \omega_{3} \omega_{4} \omega_{5}
- 128 \omega_{2}^3 \omega_{4}^3 
- 135 \omega_{2}^2 \omega_{3}^3 \omega_{5} 
+ 36 \omega_{2}^2 \omega_{3}^2 \omega_{4}^2 \right) \,, \\
&& P := 24 \left( 5 \omega_{3} \omega_{5} - 2 \omega_{4}^2 \right) \,, \\
&& R := 16 \left( 25 \omega_{2} \omega_{5}^2 
- 15 \omega_{3} \omega_{4} \omega_{5} 
+ 4 \omega_{4}^3 \right) \,, \\
&& \Delta_{0} := 60 \omega_{1} \omega_{5} - 24 \omega_{2} \omega_{4}
+ 9 \omega_{3}^2 \,, \\
&& S := 16 \left( 500 \omega_{1} \omega_{5}^3
- 200 \omega_{2} \omega_{4} \omega_{5}^2 - 225\omega_{3}^2 \omega_{5}^2 
+ 240 \omega_{3} \omega_{4}^2 \omega_{5} - 48 \omega_{4}^4 \right) \,.
\end{eqnarray*}
Then the nature of the roots are as follows\footnote{Even though there seems to be uncovered 
cases, these cannot occur as explained in \cite{Wiki}.}: \\
\underline{i) $\Delta<0$:} There are two (distinct) real and a pair of complex conjugate roots.\\
\underline{ii) $\Delta>0$:} When $P<0$ and $S<0$, all four roots are real and distinct.\\ 
When $P>0$ or $S>0$, there are two pairs of complex conjugate roots.\\
\underline{iii) $\Delta=0$:} When $P<0$ and $S<0$ and $\Delta_{0} \neq 0$, there are one 
real double and two real simple roots.\\
When $S>0$ or ($P>0$ and ($S \neq 0$ or $R \neq 0$)), there are a real double root and a pair 
of complex conjugate roots.\\
When $\Delta_{0} = 0$ and $S \neq 0$, there are a triple root and a simple root, all real.\\
When $S=0$ and $P<0$, there are two real double roots.\\
When $S=0$ and $P>0$ and $R=0$, there are two complex conjugate double roots.\\
When $S=0$ and $\Delta_{0} = 0$, all four roots are equal to \(- \omega_{4}/(5 \omega_{5}) \).

Further defining \cite{Wiki}
\begin{eqnarray*}
&& p := \frac{3}{25 \omega_{5}^2} \left( 5 \omega_{3} \omega_{5} - 2 \omega_{4}^2 \right)  \,, \\
&& q := \frac{50 \omega_{2} \omega_{5}^2 - 30 \omega_{3} \omega_{4} \omega_{5} 
+ 8 \omega_{4}^3}{125 \omega_{5}^3} \,, \\
&& \Delta_{1} := 54 \left( -20 \omega_{1} \omega_{3} \omega_{5}
+ 8 \omega_{1} \omega_{4}^2 + 10 \omega_{2}^2 \omega_{5}
- 4 \omega_{2} \omega_{3} \omega_{4} + \omega_{3}^3 \right) \,, \\
&& Q := \sqrt[3]{\frac{\Delta_{1} + \sqrt{\Delta_{1}^{2}-4 \Delta_{0}^{3}}}{2}} \,, \\
&& T := \frac{1}{2} \sqrt{-\frac{2}{3} p + 
\frac{1}{15 \omega_{5}} \left( Q + \frac{\Delta_{0}}{Q} \right) } \,, \\
\end{eqnarray*}
where any one of the three cube roots of $Q$ can be used in $T$ \cite{Wiki},
the generic roots of the quartic polynomial $p(\lambda)$ \re{pqDele} are \cite{Wiki}
\begin{eqnarray*}
&& \lambda_{1,2} = - \frac{\omega_{4}}{5 \omega_{5}} - T 
\pm \frac{1}{2} \sqrt{- 4 T^2 - 2 p + \frac{q}{T}} \,, \\
&& \lambda_{3,4} =  - \frac{\omega_{4}}{5 \omega_{5}} + T 
\pm \frac{1}{2} \sqrt{- 4 T^2 - 2 p - \frac{q}{T}} \,.
\end{eqnarray*}
Finally the \emph{real roots} \( \lambda_{i}, \, (i=1,2,3,4), \) can be used in the 
polynomial $p(\lambda)$ \re{pqDele} to tune up the bare cosmological constant $\Lambda$.

When one sets \(\omega_{2} = \omega_{3} =  \omega_{4} = 0 \), it follows 
by examining \( \Delta = 32000 \omega_{1}^3 \omega_{5}^3 \) and 
\( S = 8000 \omega_{1} \omega_{5}^3 \) that $q(\lambda)$
has no real roots when \( \omega_{1} \omega_{5} > 0 \), i.e. when \( \omega_{5} > 0 \).
When \( \omega_{5} < 0 \), there are two distinct real roots given by
\[ \lambda_{\pm} = \pm \sqrt[4]{-\frac{\omega_{1}}{5 \omega_{5}}} \,. \]
These are also roots of $p(\lambda)$ if
\[ \Lambda(\lambda_{\pm}) = \pm 36 \omega_{1} \sqrt[4]{-\frac{\omega_{1}}{5 \omega_{5}}} \,. \]

\end{document}